\documentclass{aa}
\usepackage{graphicx}
\usepackage{txfonts}

\usepackage{lscape}

\begin{document}
\title{On the nature of Nova\,1670 (CK\,Vulpeculae): \\
a merger of a red giant with a helium white dwarf}

\author{Romuald Tylenda\inst{\ref{inst1}} \and Tomek Kami\'nski\inst{\ref{inst1}} \and Radek Smolec\inst{\ref{inst2}} }

\institute{\centering  Nicolaus Copernicus Astronomical Center, Polish Academy of Sciences, 		  
       Rabia\'{n}ska 8, 87-100 Toru\'{n}, Poland,
       \email{tylenda@ncac.torun.pl, tomkam@ncac.torun.pl} \label{inst1}
\and Nicolaus Copernicus Astronomical Center, Polish Academy of Sciences, 		  
       Bartycka 18, 00-716 Warszawa, Poland,
       \email{smolec@camk.edu.pl} \label{inst2}
       }

\date{Received; accepted}

\abstract
{Nova 1670 is a historical transient bearing strong similarities to a recently-recognized type of stellar eruptions known as red novae, which are thought to be powered by stellar mergers. The remnant of the transient, CK\,Vul, is observable today mainly through cool circumstellar gas and dust, and recombining plasma, but we have no direct view on the stellar object.}
{Within the merger hypothesis, we aim to infer the most likely makeup of the progenitor system that resulted in Nova 1670.}
{We collect and summarize the literature data on the physical properties of 
the outburst and the remnant (including their energetics), and on the chemical 
composition of the circumstellar material (including elemental and isotopic 
abundances) which resulted from optical and submillimeter observations of 
the circumstellar gas of CK\,Vul. We perform simple simulations to analyze 
the form and level of mixing of material associated with the merger. 
Products of nuclear burning are identified, among them ashes of hydrogen 
burning in the CNO cycles and the MgAl chain, as well as of partial helium burning.}
{Based primarily on the luminosity and chemical composition of the remnant, we find that the progenitor primary had to be an evolutionarily advanced red-giant branch star of a mass of 1--2\,M$_{\odot}$. The secondary was either a very similar giant, or, more likely, a helium white dwarf. While the eruption event was mainly powered by accretion, we estimate that about 12\% of total energy might have come from helium burning activated during the merger. The coalescence of a first-ascent giant with a helium white dwarf created a star with a rather unique internal structure and composition, which resemble these of early R-type carbon stars.}
{Nova 1670 was the result of a merger between a helium white dwarf and a first-ascent red giant and is likely now evolving to become an early R-type carbon star.}

\keywords{ 
	stars: individual: CK\,Vul -
	stars: individual: Nova\,1670 -
        stars: activity -
	stars: winds, outflows - 
	stars: variables: other}

\titlerunning{On the nature of CK\,Vul}
\authorrunning{Tylenda, Kami\'nski, \& Smolec}
\maketitle
       

\section{Introduction \label{intro_sec}}

On 20 June 1670, Anthelme Voituret (Dom Anthelme), 
a Carthusian monk in Dijon (France),
spotted a new star of $\sim$3\,mag in the constellation of Cygnus.
A month later, the star was independently found by Johannes Hevelius
in Gda\'nsk (Poland). Since then the
object was gradually fading and, in October, it became invisible to the naked
eye. In March 1671, the star reappeared and on 30 April attained 
its maximum brightness of $\sim$2.6\,mag. 
Subsequent fading lasted till the end of August, when the
star again became invisible. Next year, in March--May 1672, 
the object was observable for a third time,
but now as a much fainter star of $\sim$5.4\,mag.
A detailed description of historical data on Nova\,1670 can be found in
\citet{shara85}.
In the modern astronomy, Nova\,1670 is known as
CK\,Vulpeculae\footnote{The constellation of Vulpecula was introduced by
J.\,Hevelius a few years after the discovery of Nova\,1670.}.

\begin{landscape}
\begin{figure}
    \includegraphics[trim=260 0 250 0,clip=True, scale=0.34]{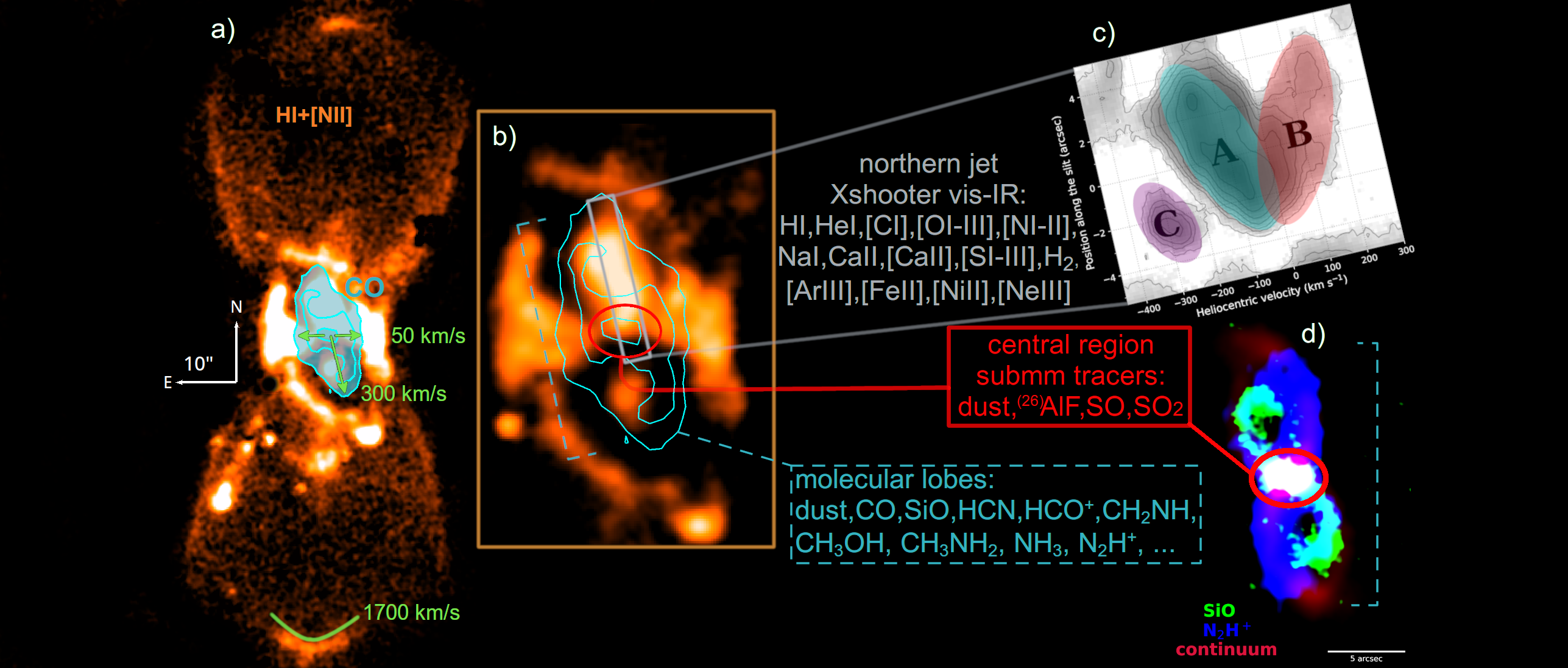}
    \caption{Components of the nebulae surrounding CK\,Vul. {\bf a)} Optical hourglass nebula (orange and white) seen in the combined emission of H$\alpha$ and [\ion{N}{II}] lines \citep[from][]{hajduk13}. Millimeter-wave emission of CO 1--0 is overlaid in cyan \citep[from][]{kmt20}. Green markers and labels show characteristic velocities of these components.
    {\bf b)} Zoomed-in central portion of the components shown in a). The structure of the optical nebula is seen more clearly. The red ellipse indicates the central region where many O-bearing molecules and strong dust emission are observed.  The gray rectangle shows the region within the northern optical ``jet'' observed spectrally in the optical--near-infrared with Xshooter at a very high sensitivity \citep{tk19}. 
    {\bf c)} The inset shows dispersed emission of a spectral line within the northern jet and illustrates that the emission is composed of at least three distinct morpho-kinematical components. 
    {\bf d)} Combined emission of SiO (green), N$_2$H$^+$ (blue), and mm-wave continuum define the molecular nebula \citep[from][]{kmt20}. The central region, bright in dust emission, is marked with a red ellipse.}
    \label{fig-collage}
\end{figure}
\end{landscape}

Numerous attempts to find a remnant of Nova\,1670, which go back to
the seventeenth century \citep[for references, see][]{shara85}, failed to detect any
stellar-like object. \citet{shara85} claimed to see 
a very faint star at the expected position, 
but later observations did not confirm that \citep{naylor}.
All what was found in the optical was a faint nebulosity emitting mainly in
the [N\,II] and H$\alpha$ lines \citep[e.g.][]{shara85,hajduk07}, shown in Fig.\,\ref{fig-collage}.

Until the end of the twentieth century, Nova\,1670 was primarily considered 
as a classical nova,  that is, an event resulting from a thermonuclear runaway on the
surface of a white dwarf (WD) accreting H-rich matter in a binary system.
This false identification of Nova\,1670 was claimed despite the obvious problem that the light curve clearly did not resemble these of classical novae (including slow novae). 
Indeed, it had two well-defined peaks separated by almost a year and followed by a much dimmer peak next year
\citep[see Fig.~2 in][]{shara85}. 
Moreover, it appeared that CK\,Vul was relatively bright in the far-infrared (IR) and submillimeter range
\citep{evans02}. With the absence of any stellar emission in the optical, this implies that the remnant of Nova\,1670 is embedded in
a thick, dusty envelope. Nothing of the sort has been observed in any of  the historical classical novae \citep[cf.][]{kamiSurv}.

These facts have prompted some authors to postulate that Nova\,1670 was the
case of a (very) late He-shell flash in a post asymptotic giant branch (post-AGB) star, similar to V605\,Aql
or V4334\,Sgr \citep{harrison,evans02}. There are, however, 
fundamental problems with
the luminosity evolution in such a scenario. Namely, Nova\,1670 had $\sim$$10^7$\,L$_\odot$ at maximum,
while CK\,Vul has only a few tens solar units at present (see
Sect.~\ref{lum_sec}), that is, the object faded by more than five orders of magnitude
during $\sim$340~years. This is to be compared to a typical luminosity of 
a post-AGB star, which is of the order of

\noindent $10^4$\,L$_\odot$ and does not change significantly during the He-shell flash.
Even if one neglects the absolute values (e.g., arguing that the distance to
CK\,Vul is uncertain), within the post-AGB 
hypothesis the luminosity drop 
would mean that CK\,Vul is below 0.1\,L$_\odot$. First, the evolution of a post-AGB star to that luminosity
would take a million years or more \citep[e.g.][]{blocker,MillerBertolami}, orders of
magnitude longer than the age of CK\,Vul. Second, it would imply a distance 
of a few tens of pc, well below any distance estimate existing in literature.
Nova\,1670 could not result from a late He-shell flash in a post-AGB star.

The eruption of V838\,Mon in 2002 \citep{munari,kimes,crause} 
was one of the most exciting and intriguing events in stellar astrophysics of a few past decades.
Astrophysicists started to realize that there exists a class of stellar outbursts 
that does not fit to any of known and understood mechanisms of stellar
eruptions. The eruptions of this new type, named red novae or red
transients, show a particular spectral evolution: in the course of the outburst
the spectrum evolves towards progressively later types, down to a late M-type
in the decline. Following the idea outlined in \citet{soktyl},
\citet{tylsok} showed that the principal characteristics of red novae can be
understood in terms of a hypothesis that these eruptions result from stellar 
mergers. Soon after, the hypothesis received strong observational support. It was mainly
presented in \citet{thk11}, who showed that V1309\,Sco was a
contact binary just before its red nova outburst in 2008.

Following the idea of \citet{soktyl}, \citet{kato} suggested that Nova\,1670
might have also resulted from a stellar merger. 
The author drew attention to
similarities between the light curve of Nova\,1670 and that of V838\,Mon.
Although the light curves of both objects display three brightness peaks,
the timescale of the outburst of Nova\,1670 is an order of magnitude longer
than that of V838\,Mon and, in fact, of any other red nova known at that time. 
This raised certain objections against the Kato's interpretation of 
Nova\,1670. However, \citet{tku13} discovered a red nova in the archives of
the OGLE project, OGLE-2002-BLG360, whose
light curve showed multiple peaks on a timescale similar to that of
Nova\,1670. This revived the interest in Nova\,1670 as a probable red nova.

Following the finding of \citet{tku13}, an extensive study of CK\,Vul in the
(sub)millimeter range was undertaken \citep{kmt15,kmt17,ktm18,kmt20}.
The project provided us with the data strongly supporting the hypothesis
of the red-nova nature of Nova\,1670. In the present paper, we summarize the
results of this project and of other studies. We then show that they can be
consistently interpreted in terms of a merger of a red-giant-branch (RGB) 
star with a helium white dwarf (HeWD).

\section{Observational results \label{obs_sec}}

\subsection{The remnant  \label{rem_sec}}

At present, the remnant of Nova\,1670 (CK\,Vul) is observed at optical and (sub)millimeter (submm) wavelengths as a bipolar nebula. Figure\,\ref{fig-collage} summarizes its components. In the optical, 
a faint emission-line nebula of an hourglass shape is visible, mainly in H$\alpha$ and [N\,II] lines \citep{shara85,hajduk07}.
The bipolar lobes extend up to $\sim$$36\arcsec$ from the hourglass center. Detailed optical spectroscopy of the brightest clump close to the center of the nebula 
revealed a rich, low-excitation nebular spectrum \citep{tk19}. No central stellar-like object has been identified so far. Evidently, it is
hidden in a dense, dusty central region of the remnant (see below).

In the submm range, the object displays a very rich spectrum of molecular emission lines \citep{kmt15,kmt17}. The emission region shows again a bipolar structure, although it is more compact than the optical nebula. The submillimeter emission of molecules and dust extend only up to $\sim$8\arcsec\ from the center \citep{kmt20}. Also, the axis of symmetry of the bipolar structure is slightly different from that of the optical hourglass. Contrary to the
optical structure, the emission region of the molecular lines (at least in some of them) and dust shows a well-defined central component of $\sim2\arcsec$ in size. It appears, that most of the molecular mass
($\sim$60\%) is concentrated in this compact central region \citep{ksb21}.

\subsection{Distance  \label{dist_sec}}

Until recently, CK\,Vul was assumed to be at a distance of $\sim$0.7\,kpc. 
This value was obtained in \citet{hajduk13} from an analysis of 
images and kinematical structure of
the nebula probed by the H$\alpha$ and [N\,II] lines. However, as pointed out in
\citet{ksb21}, \citet{hajduk13} made a mistake in their calculations. 
Instead of the inclination angle of the nebula of $i = 25\degr$ (relative to the sky plane), which resulted
from fitting their model image to the observations, 
they took $i = 65\degr$ (presumably taken as 90$\degr$ -- 25$\degr$) when deriving the distance.
If \citet{hajduk13} had taken $i = 25\degr$, they would have obtained
a distance of $\sim$3.2\,kpc \citep[see also][]{bge20}.
\citet{ksb21} showed that the inclination
cannot be larger than $\sim 35\degr$ and consequently the distance cannot be
lower than $\sim$2.6\,kpc. With their measurements of the radial velocity
difference between the outermost tips of the optical hourglass nebula, i.e.
$\sim$1640~km\,s$^{-1}$, and adopting 
$i = 25\degr$, \citet{ksb21} derived a distance of $\sim$3.5\,kpc.

The above values of the distance rely on the observed radial velocity 
difference between the northern and southern tips of the nebula
\citep[see Fig. 1 in][]{hajduk13}
and on the assumption that the line joining
the tips has the same inclination as the main body of the
hourglass nebula. The latter may not be the case, given the fact that
there are several bipolar structures in the inner nebula that are not
perfectly aligned with the axis of the large hourglass \citep{kmt15,ksb21,eez18}.

Both, the
inclination angle and the distance can be constrained from modelling the
shape of the H$\alpha$ line seen through a long slit and shown in Figs. 3
and 4 of \citet{hajduk13}. We attempted such a determination and used essentially the same model as that in
\citet{ksb21}. In particular, a linear  ("Hubble law"-like) expansion of the nebula 
was assumed  
$$\varv = \varv_0 \frac{r}{r_0},$$
where $\varv$ is the expansion velocity at a radial distance, $r$, 
from the center of the nebula. 
Simulating long-slit observations at the position shown in
Fig.~3 of \citet{hajduk13}, we were able to reproduce the observed shape 
of H$\alpha$ as
displayed in their Fig.~4. The best fit was obtained for 
$i = 24\degr$ and
$\varv_0$ = 1200\,km\,s$^{-1}$ at $r_0 = 30\arcsec$. Assuming that the
nebular expansion started 340 years ago, the obtained value of $\varv_0$ 
(at $r_0 = 30\arcsec$) results in a distance of 2.9\,kpc. Note that 
the observed radial velocity difference between the tips implies, in our
model, that the inclination of the line joining the tips is $\sim$30$\degr$.

Finally, \citet{ksb21} derived a distance of $\sim$3.8\,kpc by modelling of the kinematical structure of
molecular lobes seen in interferometric ALMA images of the molecular component of the remnant of CK\,Vul.
Henceforth, we adopt a distance to CK\,Vul of 3.3\,kpc, which is a mean value from the above determinations.

\subsection{Kinematics  \label{kin_sec}}

CK\,Vul displays a wide range of outflow velocities. The central
region expands relatively slowly, with a typical velocity of 
$\sim$50\,km\,s$^{-1}$ \citep{ksb21}. 
This is the reason why this region remains compact
compared to the rest of the remnant.
The molecular matter in the lobes expands much faster, with a characteristic
velocity of $\sim$300\,km\,s$^{-1}$ \citep{ksb21}. Even faster outflows are
observed in the optical hourglass. According to our analysis described in
Sect.~\ref{dist_sec}, the matter situated $30\arcsec$ from the center of the
remnant expands at 1200\,km\,s$^{-1}$, while the maximum outflow velocity of 
$\sim$1700~km\,s$^{-1}$ is attained in the hourglass tips (cf. panel a in Fig.\,\ref{fig-collage}). All the
values quoted in this section follow quite closely the linear expansion relation,
i.e. $\varv = $1200\,km\,s$^{-1}(r/30\arcsec)$, which is consistent with
saying that all the matter observed now in CK\,Vul was ejected more or less
at the same time, that is, during the outbursts observed in 1670--72.

\subsection{Luminosities and energetics  \label{lum_sec}}

Nova\,1670 reached its maximum visual brightness at $m_\varv \simeq 2.6$ on
30~April~1671 \citep{shara85}. 
Adopting a distance of 3.3~kpc and an interstellar extinction with 
$A_\varv \simeq 2.8$ \citep{tk19}, yields an absolute magnitude  
$M_\varv \simeq -12.8$. This
corresponds to a luminosity of $\sim 1.0\times10^7$\,L$_\odot$, if no bolometric
correction (BC = 0.0) is assumed. The power of the outburst of CK\,Vul by far exceeds 
that attained by the brightest Galactic red nova, V838~Mon
($M_\varv = -9.8$ and $L \simeq 1.0\times10^6$\,L$_\odot$) in its 2002 outburst \citep{sparks}.

The visual light curve of Nova\,1670 \citep{shara85} yields an
estimate of the total energy emitted by the object as optical radiation
of $\sim 3.3\times10^{47}$\,erg. This is rather a lower limit to the total 
radiative energy of the nova, as we do not know when the eruption started
\citep[the object was already as bright as $\sim$3.0\,mag, when it was discovered;
see][]{shara85}. Moreover, as generally red novae show, the object was likely
to emit a significant flux in the infrared in late stages of the eruption.
The above value can be compared to the kinetic energy of
the molecular ejecta of $\sim\!2.3\times10^{47}$\,erg estimated in \citet{ksb21}.

At present, CK\,Vul radiates mainly in the far-IR, with a maximum  
flux ($\lambda F_{\lambda}$) at 100--200~$\mu$m \citep{evans02,kmt15}.
Readily, the central object is obscured, and we can only observe the light re-radiated by dust.
A combination of the observed integrated flux with a distance of 3.3~kpc results in 
a luminosity of 
$\sim$20~L$_\odot$. This is a lower limit to the luminosity of the
central star, as the observed dust is clearly concentrated in a compact
torus or disc \citep{eez18,kmt20} and a certain part of the stellar radiation 
can escape through polar directions without interacting with dust. However,
the observed distribution of dust emission suggests that 
the escaping part cannot be much higher 
than the observed one \citep[cf.][]{eez18}. We thus put a value of 
100~L$_\odot$ as a generous
upper limit on the luminosity of the central object in CK\,Vul.

\subsection{Elemental abundances and isotopic ratios  \label{abund_sec}}

\begin{table}
\caption{Elemental abundances and isotopic ratios.}\label{abund_tab}
\centering
\begin{tabular}{ c c c }
\hline
  Element  & CK\,Vul & Sun \\
\hline
H  &   0.49$\pm$0.14  &   0.74 \\
He &   0.51$\pm$0.14  &   0.25 \\
C  &   ($\sim7\times10^{-4}$)\tablefootmark{a} & $2.4\times10^{-3}$ \\
N  &   $(1.2\pm0.4)\times10^{-3}$ & $7.0\times10^{-4}$ \\
O  &   $(9.5\pm3.8)\times10^{-4}$ & $5.8\times10^{-3}$ \\
Ne &   $(7.3\pm2.5)\times10^{-4}$ & $1.3\times10^{-3}$ \\
S  &   $(8.9\pm1.4)\times10^{-5}$ & $3.1\times10^{-4}$ \\
Ar &   $(2.4\pm0.6)\times10^{-5}$ & $6.7\times10^{-5}$ \\
  \\
$^{12}$C/$^{13}$C  & $3.8\pm1.0$  &  90.  \\
$^{14}$N/$^{15}$N  & $20.\pm10.$   & 440.  \\
$^{16}$O/$^{17}$O  & $\gtrsim$180.\tablefootmark{b}  & 2600.  \\
$^{16}$O/$^{18}$O  & $36.\pm14.$  &  500.  \\
$^{27}$Al/$^{26}$Al & $7.\pm3.$ & $\infty$  \\
\hline
\end{tabular}
\tablefoot{Elemental abundances are given as mass fractions. 
\tablefoottext{a}{The mass abundance of C is estimated assuming the number ratio of C to O of about 1, as suggested by the molecular composition of the gas.}
\tablefoottext{a}{Lower limit.}}
\end{table}

Table \ref{abund_tab} presents the elemental abundances and the isotopic
ratios measured in CK\,Vul. The elemental abundances were determined by 
\citet{tk19} from their spectroscopic optical
observations of the brightest nebular region of CK\,Vul (cf. Fig.\,\ref{fluor_sec} a and b). 
The abundances given in
Table\,\ref{abund_tab} are taken from 
Table 6 of \citet{tk19}.
Note that the abundance of carbon was not 
directly measured. From their study of the molecular spectrum of CK\,Vul, 
\citet{kmt15,kmt17} concluded that the C/O number ratio is close to
one. This was inferred from the inventory of molecules compared to molecular species observed around stars with known C to O ratio. Thus, the carbon abundance given in Table\,\ref{abund_tab} 
was calculated from that of oxygen, assuming C/O = 1 (in number of atoms).
The isotopic ratios of the CNO elements are taken from \citet{kmt17}, 
who performed an analysis of the rich molecular spectrum of CK\,Vul observed in
the millimeter and submillimeter region. 
The ratio of aluminum isotopes was measured by \citet{ktm18}, who carried
out observations and analysis of the emission lines of AlF in
the millimeter-wave spectrum of CK\,Vul. This was the first detection of an
astrophysical molecule containing a radioactive isotope ($^{26}$Al). 

The abundance data obtained for CK\,Vul are compared in
Table\,\ref{abund_tab} to solar values from \citet{asplund}.
Helium and nitrogen are overabundant, while oxygen is clearly underabundant in CK\,Vul.
As a result, the N/O number ratio is equal to $\sim$1.4 and is $\sim$10 times higher 
than in the Sun. All this clearly evidences that the observed
matter was processed in the CNO cycles of H burning. The heavier elements, 
which are
not affected by the CNO cycles -- that is, Ne, S, and Ar -- are, on
average, by a factor of $\sim$2.7 less abundant than in the Sun. Also,
the summed up abundance of C, N, and O -- which is not affected by the CNO cycles
either -- is lower by a similar factor, $\sim$3.2. 
Readily, the progenitor of CK\,Vul was formed in an environment $\sim$3 times 
less enriched
in heavy elements than the Sun ([Fe/H] $\simeq$ --0.5).

There are observational hints that the element abundances are not
uniform across the ejecta of CK\,Vul. An analysis of emission maps of 
different molecules in \citet{kmt20} suggests that the central
region (cf. Fig.\,\ref{fig-collage}) is more oxygen rich than the bipolar lobes, while the latter are most probably
nitrogen-rich and have the C/O number ratio close to one. The bright nebular
region investigated by \citet{tk19} overlaps spatially with the northern molecular lobe. 
The bright and extended optical components A and B 
defined in \citet{tk19} and represented in panel c in Fig.\,\ref{fig-collage} are most certainly physically linked to that lobe. However, the position of the faint and compact
component C suggests that it might belong to the central region, instead.
Although the abundance analysis of component C is less certain (due to a faint and
lower-quality spectrum) than for components A and B, the derived N/O ratio for component C is 
about four times lower than that obtained for A and B \citep[see
Table 5 in][]{tk19}. Thus, we can conclude that the matter in the lobes is
probably more affected by CNO cycles than that in the central region.

Aside to the discussed elements, an optical absorption line of lithium was observed in the outflow of CK\,Vul against more distant stars \citep{hajduk13}. Although, \citet{lit} made an attempt to derive the relative Li abundance and obtained $\log N_{\rm Li} / N_{\rm tot} + 12 \approx 2.5$ (mass fraction 2$\times 10^{-9}$), the results are very uncertain and not particularly constraining for this study. Lithium abundance and its origin are therefore not discussed here in more detail.

\subsection{Masses involved in the molecular envelope  \label{mass_sec}}

\citet{kmt17} determined parameters of the molecular matter in CK\,Vul from
their single-dish submm observations. They
derived the size of the CO emission region, $\phi$ = 9\farcs6, and the surface CO density, 
$\Sigma$(CO) = $3.0\times10^{17}$~cm$^{-2}$. 
With the adopted distance to
CK\,Vul ($d$=3.3~kpc) we obtain the total number of the CO molecules,
$N$(CO) = $\pi$\,($0.5\phi$\,$d$)$^2$\,$\Sigma$(CO) 
$\simeq 5.1\times10^{52}$ 
and the mass of CO of $\sim$$2.4\times10^{30}$\,g or $\sim$$1.2\times10^{-3}$\,M$_\odot$.
With the mass fraction of
oxygen, $\sim \! 1.0\times10^{-3}$ (Table\,\ref{abund_tab}), as well as 
assuming C/O~$\simeq$~1 and that most 
of carbon and oxygen is locked in CO, we estimate that the total 
mass of the molecular envelope of CK\,Vul is $\sim$0.7\,M$_\odot$. A similar
estimate, $\sim$0.6\,M$_\odot$, has been given by \citet{ksb21} based
on interferometric ALMA maps of CK\,Vul. However, as these authors warn, the total mass
can be even higher than $\sim$0.8\,M$_\odot$ when corrected for various
observational effects.
According to \citet{ksb21}, most of the mass ($\sim$60\%) of the circumstellar matter 
is concentrated in the central region, while the lobes contain the remaining $\sim$40\% of the molecular
material.

On the basis of the observed lines of AlF, \citet{ktm18} concluded
that the ratio of $^{27}$Al to $^{26}$Al in CK\,Vul is $\sim$7 and that the
observed mass of $^{26}$Al is $\sim \! 7\times10^{25}$\,g
(3.5$\times10^{-8}$\,M$_\odot$)
(recalculated to a distance of 3.3~kpc).  
However, the latter
value is rather a lower limit on the true amount of $^{26}$Al in CK\,Vul, 
since the abundance of AlF is limited by the
abundance of fluorine rather than that of the much more abundant aluminum.

We estimate the expected total mass of aluminum in the envelope of CK\,Vul from
its presumed abundance. As discussed in
Sect.~\ref{abund_sec}, heavy elements
in CK\,Vul are underabundant by a factor of $\sim$3 
compared to the solar composition.
We assume that a similar factor holds for aluminum and that 
the mass fraction of this element is $\sim1.9\times10^{-5}$. Then, we
expect $\sim2.6\times10^{28}$\,g ($1.31 \times 10^{-5}$ M$_{\sun}$) of aluminum in the envelope of CK\,Vul.
If the ratio $^{27}$Al/$^{26}$Al\,$\simeq$\,7 holds for the whole
envelope, the total mass of $^{26}$Al should be $\sim \! 3.7\times10^{27}$\,g
(1.85$\times10^{-6}$\,M$_\odot$).
This should be considered as an upper limit, since the
ratio of $^{27}$Al/$^{26}$Al in outer parts of the envelope is likely to be 
much higher than the value derived by \citet{ktm18} for the innermost regions.

\subsection{Abundance of fluorine  \label{fluor_sec}}

Fluorine is observed in CK\,Vul only through rotational transitions of the AlF molecule \citep{kmt17}.
Detailed ALMA images \citep{kmt20} show that the emission of AlF is
concentrated in the inner central region of the object. From the 
surface densities of AlF and CO in the central region, as derived in
\citet[][see their Fig. 8]{kmt20}, we can deduce the number ratio
F/O\,$\simeq$\,$2.2\times10^{-4}$. Here, we assume that most of the oxygen
is locked in CO and most of the fluorine is locked in AlF. The former
assumption is probably correct, given that C/O\,$\simeq$\,1 in the envelope of CK\,Vul. However, it
is questionable whether we observe all the fluorine via AlF lines, as the
existence of other molecules bearing this element, for example HF, is not excluded.
Thus, the above value of $2.2\times10^{-4}$ is most probably a lower limit 
to the F/O
ratio in CK\,Vul. Note that the solar value of F/O is $7.4\times10^{-5}$.

\section{Interpretation  \label{interpret_sec}}

In this section, we present an interpretation of the observational results
discussed so far in the framework of the hypothesis 
that the Nova\,1670 event resulted from a merger
of a binary system. We are going to constrain parameters and the
evolutionary status of the stars before the merger, as well as the nature
of physical processes that occurred during the merger event. We
start with constraining the primary component by identifying the star that was
responsible for the principal characteristics of the observed chemical
composition of CK\,Vul.

\subsection{The nature of the progenitor primary  \label{primary_sec}}

The low luminosity of CK\,Vul, 
20\,L$_\odot$\,<\,$L$\,<\,100\,L$_\odot$ 
(see Sect.~\ref{lum_sec}), indicates that the central
object is not a massive star.
Stars more massive than $\sim$3~M$_\odot$ are more luminous than 100\,L$_\odot$ at
any evolutionary stage \citep{Iben,Karakas}. The primary (and the secondary) 
of the progenitor binary could not have been more massive than that.
We can also exclude an AGB star of any mass as the primary (or secondary) of the
progenitor. The dense CO-core would have easily survived 
a merger with almost any
other star and the remnant would now be at least as luminous as the progenitor,
i.e. >10$^3$\,L$_\odot$.

The observed abundances in CK\,Vul (overabundance of helium and nitrogen,
underabundance of oxygen -- see Table~\ref{abund_tab}) evidently show that
some part of the observed matter has been processed in the CNO cycles
of H burning. This and the presence of $^{26}$Al
strongly favor the red-giant-branch (RGB) phase of the progenitor primary 
(if a massive or AGB star is excluded, see above). 
Moreover, the abundance of
$^{26}$Al shows that it had to be an evolutionarily advanced RGB star, since the isotope of $^{26}$Al is effectively
produced in low-mass RGB stars only when the He-core
mass is $\gtrsim$0.2~M$_\odot$.

The only way to obtain $^{26}$Al is to activate the MgAl chain, 
which includes the $^{25}$Mg(p,$\gamma$)$^{26}$Al reaction.
This takes place when the burning
temperature exceeds $3.0\times10^7$\,K \citep[e.g.,][]{arnould}.
Then, $^{25}$Mg is converted to $^{26}$Al in the
H-burning shell. $^{26}$Al is unstable and decays through
$^{26}$Al($\beta^+$)$^{26}$Mg on a ($e$-folding) timescale
$\tau_{\beta^+} = 1.03\times10^6$ yr \citep{halflife}.

We can estimate the luminosity of the presumed RGB star from
the observed mass of $^{26}$Al.
The luminosity of an RGB star, $L$,
is related to the rate of burning hydrogen in the star, 
$\zeta(\rm{H})$, through
$$L = \epsilon_{\rm{CNO}}\,\zeta(\rm{H}),$$
where $\epsilon_{\rm{CNO}}$ is energy obtained from burning a unit
mass of hydrogen in the CNO cycles 
($\epsilon_{\rm{CNO}} = 6.0\times10^{18}$~erg\,g$^{-1}$).

Let us assume that all $^{25}$Mg existing 
in the unprocessed matter is converted to $^{26}$Al in the H-burning shell.
The rate of this process is related to the speed of 
the hydrogen burning, so we can write
$$\zeta(^{25}{\rm Mg}) =  \zeta({\rm H})\frac{X(^{25}{\rm Mg})}{X({\rm H})},$$
where $X(^{25}$Mg) and $X$(H) are the mass fractions of $^{25}$Mg and H,
respectively, prior to the H burning. Let us further assume that the distribution of $^{26}$Al in
the He core is governed by the production of this isotope in the H-burning
shell and its subsequent $\beta^+$ decay in the core. 
The rate of production of
$^{26}$Al, $\xi(^{26}\rm{Al})$, is equal to the rate of destruction of
$^{25}$Mg, $\zeta(^{25}\rm{Mg})$. Then the total mass of $^{26}$Al 
in the core, $M(^{26}$Al), is given by
$$M(^{26}\rm{Al}) = \xi(^{26}\rm{Al})~\tau_{\beta^+}.$$
After a simple transformation of the above formulae, we get
$$L = \frac{\epsilon_{\rm{CNO}}}{\tau_{\beta^+}}\,
\frac{X\rm{(H)}}{X(^{25}\rm{Mg})}\,M(^{26}\rm{Al}).$$
Assuming that the initial abundance of magnesium in the CK\,Vul
progenitor was 3 times lower than that in the Sun and that 10\% of it 
was in the form of $^{25}$Mg \citep[like in the Sun
-- see][]{asplund}, we get $X$(H)/$X(^{25}$Mg) = $3.0\times10^4$. 
Then the above formula can be rewritten as
$$L = 285\,\rm{L}_\odot\,
\frac{M(^{26}\rm{Al})}{1.0\times10^{-7}\,\rm{M}_\odot}.$$
Substituting the upper and lower limits on the observed mass of 
$^{26}\rm{Al}$ derived in Sect.~\ref{mass_sec}, i.e. 
3.5$\times10^{-8}$\,M$_\odot<M(^{26}\rm{Al})<1.85\times10^{-6}$\,M$_\odot$,
to the above formula, we obtain limits on the luminosity of the CK\,Vul
progenitor, namely
$$100\,{\rm L}_{\odot}<L<5300\,{\rm L}_\odot.$$
These are fairly consistent with the values expected for low-mass stars in
an advanced RGB phase, that is, those having the He-core mass of 0.3--0.45\,M$_\odot$
and the corresponding luminosity of 100--2000\,L$_\odot$ \citep[cf.][]{Spada,Hildago}.

The lack of optical detection of the very highly dust-embedded product of the merger can also be used to infer the properties of the progenitor. \cite{Morgan22} considered dust condensation in systems undergoing a merger and found relations between accumulated dusty material and the ``compactness'' of the primary progenitor, namely between the mass and radius of the progenitor and the circumstellar extinction in the orbital plane, parametrized by $A_V$. Using their relation, we can constrain $A_V$ in CK\,Vul. In $R$-band observations of the field of CK\,Vul of \citet{hajduk13}, the weakest detected stars have $\sim$22.5 mag. Adopting again the interstellar extinction of $A_R \approx A_V = 2.8$ mag, we take 25.3 mag as a lower limit on the $R$-band magnitude of the star. Then, to hide any star with a luminosity >20\,L$_{\odot}$ (Sect.\,\ref{lum_sec}), the dusty torus surrounding CK\,Vul should cause reddening with $A_V \gtrsim 11$\,mag. In this derivation, we considered stellar colors and absolute magnitudes of giants, red supergiants, and main-sequence stars of intermediate spectral types. Reinstating our argument from the beginning of this section, that the progenitor had to be of a mass lower than $\sim$3\,M$_{\sun}$, we find from \citet{Morgan22} that to form a thick torus with $A_V \gtrsim 11$\,mag, the primary progenitor had to be large, with a radius consistent -- for a given mass -- with an RGB phase; for instance, we get a radius >40\,R$_{\sun}$ for a 1\,M$_{\sun}$ star.

We conclude that the primary component of the CK\,Vul progenitor
binary was most probably a low-mass (initial mass of $\sim$1--2~M$_\odot$)
star in the RGB phase.

\begin{table*}
\begin{center}\small
\caption{Simulated elemental abundances and isotopic ratios 
(see text) compared to ones observed in CK\,Vul.}
\label{model_tab}
\begin{tabular}{ c c c c c c c c}
\hline
  element &  H-envelope  &  He-core &   mixed  &   CK\,Vul &   mixed &  H-envelope  &  He-core\\
  &  & (outer 0.1 M$_\odot$) &  $f_{\rm He}$=0.40& & $f_{\rm He}$=0.40&&(outer 0.1 M$_\odot$)\\
\hline
&\multicolumn{3}{c}{\bf integrated Tycho mass fraction}&observed&\multicolumn{3}{c}{\bf integrated MESA mass fraction}\\ \cline{2-4}\cline{6-8}\\[-5pt]
H\tablefootmark{a} &        0.74   &    0.00   &    0.44    &   0.49 & 0.43	& 0.72 & 0.00\\
He\tablefootmark{a}&        0.25    &   1.00   &    0.55   &    0.51 & 0.56	& 0.28 & 1.00\\
\\[-5pt]
C&$7.6\times10^{-4}$&$6.0\times10^{-5}$&$4.8\times10^{-4}$&$(7\times10^{-4}$)\tablefootmark{b} &3.8$\times 10^{-4}$&5.9$\times 10^{-4}$&5.6$\times 10^{-5}$\\
N\tablefootmark{a}&$2.2\times10^{-4}$&$2.6\times10^{-3}$&$1.2\times10^{-3}$&$1.2\times10^{-3}$ &1.2$\times 10^{-3}$&3.2$\times 10^{-4}$&2.6$\times 10^{-3}$\\
O\tablefootmark{a}&$1.8\times10^{-3}$&$6.1\times10^{-5}$&$1.1\times10^{-3}$&$9.5\times10^{-4}$ &1.1$\times 10^{-3}$&1.9$\times 10^{-3}$&3.3$\times 10^{-5}$\\
F&$1.6\times10^{-7}$&$2.0\times10^{-10}$&$9.5\times10^{-8}$&(>$2.3\times10^{-7}$)\tablefootmark{c}&3.3$\times 10^{-10}$&5.6$\times 10^{-10}$&3.1$\times 10^{-12}$\\
Ne\tablefootmark{a}&$4.0\times10^{-4}$&$3.4\times10^{-4}$&$3.8\times10^{-4}$&$7.3\times10^{-4}$&3.9$\times 10^{-4}$&4.2$\times 10^{-4}$&3.4$\times 10^{-4}$\\
S&$9.8\times10^{-5}$&$9.8\times10^{-5}$&$9.8\times10^{-5}$&$8.9\times10^{-5}$&\\
Ar&$2.1\times10^{-5}$&$2.1\times10^{-5}$&$2.1\times10^{-5}$&$2.4\times10^{-5}$&\\
 \\
&\multicolumn{3}{c}{\bf integrated Tycho number ratio}&observed&\multicolumn{3}{c}{\bf integrated MESA number ratio}\\ \cline{2-4}\cline{6-8}\\[-5pt]
   C/O    &       0.55   &    1.28  &     0.57   &  $\sim$1\tablefootmark{b}  &0.86	&0.31	&1.68\\
   N/O    &       0.14   &   48.6   &    1.19  &     1.38                     &31.29	&0.17	&77.97\\
F/O&$7.4\times10^{-5}$&$2.7\times10^{-6}$&$7.3\times10^{-5}$&(>$2\times10^{-4}$)\tablefootmark{c}&2.2$\times 10^{-7}$ &	3.0$\times 10^{-7}$ & 9.6$\times 10^{-8}$\\
\\[-5pt]
$^{12}$C/$^{13}$C  & 99. & 3.5  &  48.  &  3.8                 &19.8	&30.2	  &4.3\\
$^{14}$N/$^{15}$N  & 470.  & 28000. & 3700.  & 20.0            &9554.	&457.	  &23200.\\
$^{16}$O/$^{17}$O  & 2780. &  9.6  &  410.0  &  $\gtrsim$180.  &1803.	&2338.	  &999.\\
$^{16}$O/$^{18}$O  & 590.  & 26000. &  600.0 &  36.0           &5.1$\times 10^{5}$	&500.	  &1.3$\times 10^{6}$\\
$^{27}$Al/$^{26}$Al & $\infty$ & 2.4&6.0 &7.0 \\
\hline
\end{tabular}
\end{center} 
\tablefoot{\tablefoottext{a}{Quantity included in the mixture fit.}\tablefoottext{b}{C/O number ratio was assumed (see text).}\tablefoottext{c}{A rough lower limit; see Sect.\,\ref{fluor_sec}.}}
\end{table*}

\subsection{Simulation of the observed abundances  \label{simul_sec}}

The abundances in CK\,Vul in Table~\ref{abund_tab} suggest that 
we observe the dispersed matter that was dredged up from the outer layers of 
the He core of the RGB primary and that was mixed with the unprocessed matter 
from the envelope. In order to show that it is very likely the case, 
we performed simple simulations.
Using the Tycho-6.0 \citep{tycho,tychocode} and MESA \citep[Modules for Experiments in Stellar Astrophysics; release 21.12.1][]{Paxton2011, Paxton2013, Paxton2015, Paxton2018, Paxton2019, Jermyn2023}, we followed the evolution of 
a $1.0\,$M$_\odot$ star up to the tip of the RGB\footnote{The MESA inlist file is available under DOI 10.5281/zenodo.10027069 and \url{https://doi.org/10.5281/zenodo.10027069}}. Solar abundances
\citep{asplund} were adopted in the initial model, but elements heavier than He were depleted by 
a factor of 3 in Tycho ([Fe/H] = --0.5); equivalently, Z=0.008 was adopted in MESA. We then took the
stellar model at the time when its He core had a mass of $0.40$\,M$_\odot$
and mixed the outer $0.1\,$M$_\odot$ region of the
He core with the unprocessed envelope material. The mixed
abundances were compared to those of CK\,Vul. We tried to obtain the best fit by
changing the fraction of the He-core
material in the final mixture, $f_{\rm He}$. The mass fractions of the elements
whose abundances were determined from the observations, that is, H, He, N, O, 
and Ne (those of S and Ar remain unchanged in the H-burning
processes), were taken into account in the fitting procedure in which 
we searched for the lowest sum of relative squared deviations between the simulated ($X_{n, {\rm sim}}$) and observed ($X_{n, {\rm obs}}$) abundances, i.e.
$$\chi^2 (f_{\rm He})=\sum_{n={\rm H,~He,~O,~Ne}} \frac{(X_{n, {\rm obs}}-X_{n, {\rm sim}})^2}{X_{n, {\rm obs}}^2}.$$
The best fit was obtained for $f_{\rm He}$=0.40 using both Tycho and MESA abundances. The results are compared in Table \ref{model_tab}. Except for F which was not considered in the fit, the elemental abundances are fairly consistent between the two evolution packages used. 

As can be seen from Table \ref{model_tab}, the abundances of the five considered
elements are reproduced quite well in our simulations. 
(To save time, S, Ar, and Al abundances were not considered in MESA.)
The relative mixed abundances of the CNO elements and the $^{27}$Al/$^{26}$Al ratio are also relatively well reproduced, especially by Tycho calculations. This was achieved by adjusting only one parameter. 
The success of this simple model can be regarded
as a strong argument in favor of our hypothesis that
the matter observed in CK\,Vul comes from the He core and the H-rich envelope
of an advanced RGB star.

There are, however, inconsistencies between the model results and
the observed values. 
Of first concern are the isotopic ratios of the CNO elements, especially if MESA results are considered becuse MESA predicts much lower abundances of $^{17}$O and $^{18}$O than Tycho.
Also, the abundances of C, F, and Ne are slightly below the observed values. 
Clearly, some part of the observed
matter had to be also processed in nuclear reactions other than
just the equilibrium H burning. We discuss this point further in Sect.~\ref{burning_sec}.

\subsection{The nature of the progenitor secondary  \label{secondary_sec}}

Within our hypothesis, the required mixing of the matter from the He core with 
that from the H-rich envelope is caused by the merger of two stars. In other
words, the merger event has to disrupt a large portion of the
He core. In order for this to happen, the secondary has to have a core that is at least as dense as
that of the primary. This rather rules out a main-sequence (MS) star as the secondary.
At the base of the layer rich in $^{26}$Al, the density of the RGB He core 
is two orders of magnitude higher than that in the center of
lowest-mass MS stars \citep[e.g.,][]{Iben74}. Thus, the He core of an advanced RGB star would survive
the merger with a low-mass MS star, remaining practically intact. 
Consequently, any dredge up of matter
from within the He core would be impossible. Moreover, the remnant of such an MS-RGB merger
would remain an RGB star with a luminosity of 
$10^2-10^3$~L$_\odot$, which is significantly higher than that of CK\,Vul.

\begin{figure}
  \includegraphics[width=\columnwidth]{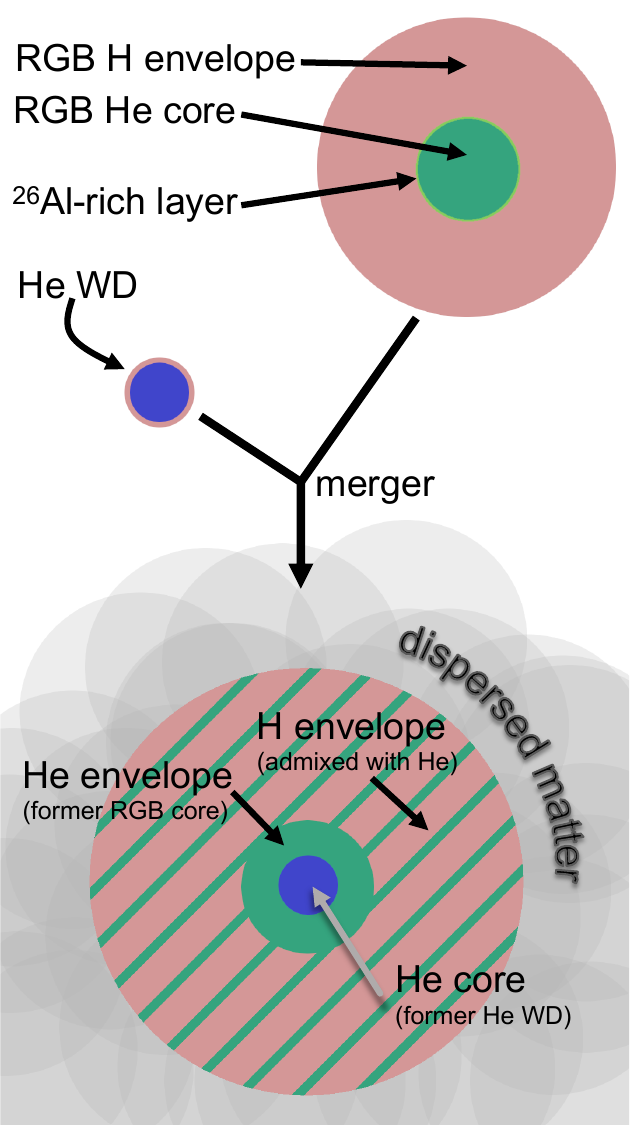}
  \caption{Schematic illustration of the proposed scenario of a merger between an RGB star with a HeWD, and of its product. The relative sizes of the internal substructures of the stars are not to scale. Pink regions show H-rich matter; dark green represents He-rich gas; light green shows the thin $^{26}$Al layer; and blue depicts the He core of the WD.}
\label{fig}
\end{figure}

Instead, a helium white dwarf (HeWD) would meet the observational requirements 
as the secondary of the binary progenitor of CK\,Vul. 
A HeWD can be significantly more compact than an RGB
He core of a similar mass \citep[e.g.,][]{Hall}. In other words, a He WD can be even several times
denser than the RGB core \citep[e.g.,][]{Iben74}. 
Therefore, in the case of a merger of an RGB star with a HeWD,
the He RGB core would become disrupted and finally would form a
He-rich envelope around the WD. Since originally the He core was
surrounded by a H-rich envelope, it is very likely that the matter from the outer
layers of the He core would be mixed with that from the H envelope \citep[for a fuller discussion of the possible scenarios, see][]{zj13}. The merger
event releases a lot of energy, therefore some part of
the mixed matter would gain enough energy to be dispersed around the system \citep[e.g.,][]{Piersanti}.
After a certain time, it could be observed as a dusty and molecular circumstellar 
envelope of the remnant. The proposed concepts are schematically illustrated in Fig.\,\ref{fig}.  

The existence of a HeWD in the progenitor binary indicates that the two stars interacted in the past. It is therefore consistent with the progenitor being a close system prone to instabilities leading to a merger. Independent of that, \citet{ksb21} speculated that the merger binary had a companion which is currently interacting with the merger ejecta. If confirmed, that would make CK Vul progenitor a triple star.

\subsection{Nuclear burning during the merger  \label{burning_sec}}

The merger of an RGB star with a HeWD can also explain the inconsistencies
between our modelling and the observations summarized in
Table~\ref{model_tab}, that is, the low values of the observed isotopic ratios
of the CNO elements and enhanced abundances of C, F, and Ne.

Let us consider the $^{16}$O/$^{18}$O ratio, which is
unexpectedly low in CK\,Vul. The only way to produce a significant amount 
$^{18}$O is to activate He burning. Then, $^{14}$N, which is the second
most abundant isotope after $^4$He in the matter that has been processed
in the CNO cycles, is converted to $^{18}$O through
$^{14}$N($\alpha$,$\gamma$)$^{18}$F($\beta^+$)$^{18}$O. However,
at the end of He burning, $^{18}$O is converted to $^{22}$Ne 
\citep[see Fig.~10 in][]{arnould}. Therefore, an enhanced abundance of
$^{18}$O can be taken as evidence of partial
He burning \citep[e.g.][]{clay07}.

Helium burning generally requires temperatures above
$\sim$1.0$\times10^8$\,K. Mergers of a HeWD with an RGB He core
can reach it \citep{zj13}. Objects of this kind and of a mass of 
0.3--0.4~M$_\odot$, when approaching each other, attain orbital velocities of
$\sim$2000\,km\,s$^{-1}$. Parcels of He-rich matter, when
colliding at these velocities, may attain temperatures of up to
$\sim \! 2.0\times10^8$\,K. Following models of mergers of 
a HeWD with a CO WD \citep[e.g.][]{long11,staff12}, 
we can thus expect in our case that after disruption of the RGB He core, a short-lived, 
hot He-rich layer is formed above the HeWD, whereby partial He burning can occur and $^{18}$O can be effectively produced. 
Once He burning is activated, we can also expect
increased abundances of $^{12}$C, $^{19}$F and $^{22}$Ne 
\citep[see e.g. Figs.~10 and 11 in][]{arnould}. 
This is thus a viable scenario to
explain the observed overabundance of C, F, and Ne in Table~\ref{model_tab}.
The case of CK\,Vul is a strong observational hint that at least some HeWD-RGB mergers activate He burning, despite earlier controversies on the topic \citep[see][and references therein]{Piersanti}. 

Stellar mergers are accompanied by strong mixing processes. In this way,
matter from stellar interiors can be observed in the outermost regions of
the merger product. In our case, it can be also expected
that parcels of H-rich gas from the RGB envelope are
admixed to the hot He-rich matter. This would activate, 
perhaps for a short time, initial reactions of the CNO cycles, leading to 
an enhancement of $^{13}$C and, possibly, of $^{17}$O in the merger 
remnant. The enhanced abundance of $^{18}$O in the He-rich matter (after
having suffered from partial He burning) would also allow production of
$^{15}$N via $^{18}$O($p$,$\alpha$)$^{15}$N.

\subsection{Contribution of He-burning to the energetics  \label{He-burn_sec}}

As mentioned in Sect.~\ref{abund_sec}, the C/O number ratio is probably close
to one in CK\,Vul, which gives the C mass fraction $\sim \! 7\times10^{-4}$ (see
Table~\ref{abund_tab}). The simple model presented
in Sect.~\ref{simul_sec} 
is not able to meet this figure (see
Table\,\ref{model_tab}). However, within our hypothesis of partial
He burning, we can get additional carbon.
Converting $\sim \! 4\times10^{-4}$ of
the observed helium to carbon would be enough to get the desired C
abundance. Burning $\sim \! 1\times10^{29}$\,g (5$\times 10^{-4}$ M$_{\sun}$) of helium would output energy
of $6\times10^{46}$\,erg. 
This is $\sim$12\% of the observed energy released by
Nova\,1670 (see Sect.~\ref{lum_sec}). 
Only a rather small fraction of this He-burning energy would 
contribute to the observed radiative luminosity. Most of it 
would probably be released in
various forms of mechanical energy (overcoming the gravitational
potential, kinetic energy of the outflowing matter, low- and large-scale
turbulent motions, etc.).

\subsection{Kinematics of the merger outflow  \label{outflow_sec}}

The merger of an RGB star with a WD explains not only the observed elemental
abundances and the isotopic ratios but also the wide range of the observed
outflow velocities, i.e. from $\sim$50~km\,s$^{-1}$ in the compact equatorial
region up to $\sim$1700~km\,s$^{-1}$ in the tips of the optical hourglass 
(see Sect.~\ref{kin_sec} and Fig.~\ref{fig-collage}). At the beginning of the merger, when the WD and the
RGB core spiral in the extended common envelope, the matter is lost mainly in
the orbital plane with velocities characteristic of winds from red giants,
i.e. a few tens of km\,s$^{-1}$. This creates a dense equatorial belt of slowly
expanding matter. At the final merger, when the WD coalesces with the He-core,
the matter is lost with velocities comparable to the escape velocity of a
WD, i.e. a few thousand km\,s$^{-1}$. This fast outflow is however blocked 
in the equatorial directions by the dense slow-moving belt of the previously lost
matter. In the polar directions, the fast outflow expands
relatively freely. In the above scenario, the bipolar lobes should be
more enriched in the nuclear ashes than the equatorial belt. This is observed 
in the remnant of CK\,Vul, for instance, C/O and N/O are enhanced 
in the lobes (Sect.~\ref{abund_sec}).

\subsection{Luminosity of the remnant  \label{remnant_sec}}

At present, 350 years after the merger, the remnant of CK\,Vul is likely 
to have a dense He-core
(the former HeWD) in its center, surrounded by an extended He-rich 
layer (the former He-core of the RGB primary) and an outer H-rich
envelope (a part of the former H-rich envelope of the primary). 
The structure is schematically illustrated in Fig.\,\ref{fig}.
It is easy to show that an object like that should be practically devoid of nuclear power sources.
The He core has a too low mass to ignite He in its center.
The base of the H-rich envelope,
situated above the extended He layer, is most probably
insufficiently hot and dense to start H burning. This explains the low current-day
luminosity of CK\,Vul (<100\,L$_\odot$).

\section{Discussion}

In the previous section, we have come to a conclusion that 
the merger of an RGB giant and a HeWD can consistently
account for the main observational data
of CK\,Vul.
This scenario allows us to explain and understand the observed elemental 
abundances, isotopic ratios, kinematics in the nebula,
as well as the low luminosity of the object in the contemporary epochs.

Alternatively, however, a merger of two RGB stars could also account for the observational characteristics of CK\,Vul.
The main difference is that in the case of a double RGB merger, 
$^{26}$Al would come
from disrupted outer layers of the more massive He core and that
a major body of this core would survive the merger and form a core of the
remnant. However, we consider this scenario as less probable than the RGB-HeWD merger, 
because it would imply almost equal masses of the binary components at the zero-age main sequence (ZAMS). 
Our Tycho-6 modelling shows
that the difference in masses of the components should be $\lesssim$3\%, 
in order to satisfy the condition that the less massive star enters 
the RGB phase before the more massive one leaves it \citep[see
also][]{sss19}. 
Although little probable, the case of a double RGB star merger cannot be excluded.
All the more that a collision of two RGB stars would probably more likely result in
merging of the two cores than merging an RGB core with a HeWD. This is because
the common envelope resulting from two RGB envelopes is expected to be more massive than that from a HeWD and a red giant.

It is interesting and instructive to note that our preferred scenario  
of a merger of a HeWD with an RGB star has been proposed to explain the origin of early-R and J type stars \citep[e.g.][]{ijl07,zj13,zj20}. These are a special
type of carbon stars \citep{dominy,abia,zamora}, 
with enhanced nitrogen and lithium, a low
$^{12}$C/$^{13}$C ratio, solar O abundances, 
solar or slightly subsolar metallicities, 
and no $s$-element enrichment. Most probably, 
they are single stars \citep{Clure}. The early-R type stars have magnitudes
similar to the red-clump stars, which suggests they are 
in the core He-burning phase. J-type stars are even brighter.

It is tempting to ask whether CK\,Vul is evolving toward an early R-type star.
The answer is not straightforward. In mergers between HeWDs and RGB stars considered by \cite{zj20}, simulations show that carbon stars are produced only for high-mass ($\gtrsim$0.45 M$_{\sun}$) WDs and at metallicities higher than what we adopted for the CK\,Vul progenitor. Whether CK\,Vul remnant is becoming a He-burning carbon star similar to early-R type stars will become more clear when the nebula, especially its dense equatorial waist, 
disperses and the central star
becomes visible. Meanwhile, the low luminosity of CK\,Vul argues in favor of
it becoming an R-type star. Its presumable structure, outlined in
Sect.~\ref{remnant_sec}, is very likely to evolve toward
the core He-burning phase. 
The N overabundance in the nebula is also in favor of that, and so is
the low $^{12}$C/$^{13}$C ratio. The observational indications
that C/O\,$\simeq$\,1 in the molecular lobes and C/O<1 in the
central dense waist may suggest that the
star is going to become, or perhaps already is, a carbon star
(C/O\,>\,1). As mentioned in Sect.\,\ref{abund_sec}, lithium appears to be overabundant in the outflow of CK\,Vul \citep[see also discussion in][]{lit}, which is also in line with early-R stars characteristics.
The only observational results that make CK\,Vul appear different from the
early R-type stars are the abundances of oxygen and heavier elements. 
While R-type stars usually show the O abundances and metallicities roughly 
solar \citep{dominy,zamora}, 
in CK\,Vul oxygen and heavier elements are significantly underabundant. As to
oxygen, we may note that observational determination of the O abundance is
difficult in carbon stars. It is often estimated from scaling with the
metallicity \citep[e.g.][]{zamora}. Nevertheless, in a sample of five early R-type
stars, for which \citet{dominy} were able to reliably estimate the O
abundance, there is one star, HD\,100764, with both 
parameters close to those found for CK\,Vul, i.e. [O]\,$\simeq-0.53$ and
[Fe]\,$\simeq-0.59$ (versus [O]\,$\simeq-0.60$ and [Fe]\,$\simeq-0.31$ in
CK\,Vul).

In conclusion, it is very likely 
that the historical Nova\,1670
resulted from a merger of a HeWD and an RGB star, and that its remnant,
CK\,Vul, is now evolving toward becoming an early R-type star.



\begin{acknowledgements}
TK acknowledges funding from grant SONATA BIS no 2018/30/E/ST9/00398 from the Polish National Science Center. RS acknowledges support
from the Polish National Science Center grant SONATA BIS 2018/30/E/ST9/00598
\end{acknowledgements}


\begin{thebibliography}{}

\bibitem[Auer et al.(2009)]{halflife}
Auer, M., Wagenbach, D., Wild, E.~M., et al.\ 2009, Earth and Planetary Science Letters, 287, 453 

\bibitem[Abia \& Isern(2000)]{abia}
Abia, C. \& Isern, J. 2000, \apj, 536, 438

\bibitem[Arnett(2013)]{tychocode}
Arnett, D.\ 2013, Astrophysics Source Code Library. ascl:1303.008

\bibitem[Arnould et al.(1999)]{arnould}
Arnould, M., Goriely, S., \& Jorissen, A. 1999, \aap, 347, 572

\bibitem[Asplund et al.(2009)]{asplund}
Asplund, M., Grevesse, N., Sauval, A.J., \& Scott, P, 2009, \araa, 47, 481

\bibitem[Banerjee et al.(2020)]{bge20}
Banerjee, D. P. K., Geballe, T. R., Evans, A. et al. 2020, \apj, 904, L23

\bibitem[Bl\"ocker(1995)]{blocker}
Bl\"ocker, T., 1995, \aap, 299, 755

\bibitem[Clayton et al.(2007)]{clay07}
Clayton G. C., Geballe T. R., Herwig F., et al. 2007, \apj, 662, 1220

\bibitem[Crause et al.(2003)]{crause}
Crause, L. A., Lawson, W. A., Kilkenny, D., et al. 2003, \mnras, 341, 785

\bibitem[De Marco et al.(2011)]{marco11}
De Marco, O., Passy, J. C., Moe, M., et al. 2011, \mnras, 411, 2277

\bibitem[Dominy(1984)]{dominy}
Dominy, J. F. 1984, \apjs, 55, 27

\bibitem[Evans et al.(2002)]{evans02}
Evans, A., van Loon, J. Th., Zijlstra, A. A., et al. 2002, \mnras, 332, L35

\bibitem[Eyres et al.(2018)]{eez18}
Eyres, S. P. S., Evans, A., Zijlstra, A., et al. 2018, \mnras, 481, 4931

\bibitem[Hajduk et al.(2007)]{hajduk07}
Hajduk, M., Zijlstra, A. A., van Hoof, P. A. M., et al. 2007, \mnras, 378, 1298

\bibitem[Hajduk et al.(2013)]{hajduk13}
Hajduk, M., van Hoof, P. A. M., \& Zijlstra, A. A. 2013, \mnras, 432, 167

\bibitem[Hall \& Tout(2014)]{Hall} 
Hall, P.~D. \& Tout, C.~A.\ 2014, \mnras, 444, 3209 


\bibitem[Hidalgo et al.(2018)]{Hildago} 
Hidalgo, S.~L., Pietrinferni, A., Cassisi, S., et al.\ 2018, \apj, 856, 125 


\bibitem[Harrison(1996)]{harrison}
Harrison, T. E., 1996, \pasp, 108, 1112


\bibitem[Iben(1974)]{Iben74} 
Iben, I.\ 1974, \araa, 12, 215

\bibitem[Iben(1985)]{Iben} 
Iben, I.\ 1985, \qjras, 26, 1

\bibitem[Izzard et al.(2007)]{ijl07}
Izzard R. G., Jeffery C. S., \& Lattanzio J., 2007, \aap, 470, 661

\bibitem[Jermyn et al.(2023)]{Jermyn2023} Jermyn, A.~S., Bauer, E.~B., Schwab, J., et al.\ 2023, \apjs, 265, 15

\bibitem[Kami{\'n}ski et al.(2022)]{kamiSurv} 
Kami{\'n}ski, T., Mazurek, H.~J., Menten, K.~M., et al.\ 2022, \aap, 659, A109

\bibitem[Kami\'nski et al.(2015)]{kmt15}
Kami\'nski, T., Menten, K. M., Tylenda, R., et al. 2015, \nat, 520, 322

\bibitem[Kami\'nski et al.(2017)]{kmt17}
Kami\'nski, T., Menten, K. M., Tylenda, R., et al. 2017, \aap, 607, A78

\bibitem[Kami\'nski et al.(2018)]{ktm18}
Kami\'nski, T., Tylenda, R., Menten, K. M., et al. 2018, Nature Astronomy, 
2, 778

\bibitem[Kami{\'n}ski et al.(2023)]{lit} 
Kami{\'n}ski, T., Schmidt, M., Hajduk, M., et al.\ 2023, \aap, 672, A196 

\bibitem[Kami\'nski et al.(2018)]{kst18}
Kami\'nski, T., Steffen, W., Tylenda, R. et al. 2018, \aap, 617, A129

\bibitem[Kami\'nski et al.(2020)]{kmt20}
Kami\'nski, T., Menten, K. M., Tylenda, R. et al. 2020, \aap, 644, A59

\bibitem[Kami\'nski et al.(2021)]{ksb21}
Kami\'nski, T., Steffen, W., Bujarrabal, V. et al. 2021, \aap, 646, A1

\bibitem[Karakas \& Lattanzio(2014)]{Karakas} 
Karakas, A.~I. \& Lattanzio, J.~C.\ 2014, \pasa, 31, e030. 

\bibitem[Kato(2003)]{kato}
Kato, T., 2003, \aap, 399, 695

\bibitem[Kimeswenger et al.(2002)]{kimes}
Kimeswenger, S., Lederle, C., Schmeja, S., \& Armsdorfer, B. 2002, \mnras,
336, L43

\bibitem[Longland et al.(2011)]{long11}
Longland, R.; Lorén-Aguilar, P.; José, J. et al. 2011, \apj, 737, L34


\bibitem[MacLeod et al.(2022)]{Morgan22} 
MacLeod, M., De, K., \& Loeb, A.\ 2022, \apj, 937, 96 

\bibitem[McClure(1997)]{Clure}
McClure, R.~D.\ 1997, \pasp, 109, 256 


\bibitem[Miller Bertolami(2016)]{MillerBertolami} 
Miller Bertolami, M.~M.\ 2016, \aap, 588, A25. 

\bibitem[Munari et al.(2002)]{munari}
Munari, U., Henden, A., Kiyota, S. et al. 2002, \aap, 389, L51

\bibitem[Naylor et al.(1992)]{naylor}
Naylor, T., Charles, P. A., Mukai, K., \& Evans, A. 1992, \mnras, 258, 449

\bibitem[Nelemans et al.(2000)]{nelemans00}
Nelemans, G., Verbunt, F., Yungelson, L. R., \& Protegies Zwart, S. F.
2000, \aap, 360, 1011




\bibitem[Paxton et al.(2011)]{Paxton2011} Paxton, B., Bildsten, L., Dotter, A., et al.\ 2011, \apjs, 192, 3
\bibitem[Paxton et al.(2013)]{Paxton2013} Paxton, B., Cantiello, M., Arras, P., et al.\ 2013, \apjs, 208, 4
\bibitem[Paxton et al.(2015)]{Paxton2015} Paxton, B., Marchant, P., Schwab, J., et al.\ 2015, \apjs, 220, 15
\bibitem[Paxton et al.(2018)]{Paxton2018} Paxton, B., Schwab, J., Bauer, E.~B., et al.\ 2018, \apjs, 234, 34
\bibitem[Paxton et al.(2019)]{Paxton2019} Paxton, B., Smolec, R., Schwab, J., et al.\ 2019, \apjs, 243, 10

\bibitem[Piersanti et al.(2010)]{Piersanti} 
Piersanti, L., Cabez{\'o}n, R.~M., Zamora, O., et al.\ 2010, \aap, 522, A80 

\bibitem[Segev et al.(2019)]{sss19}
Segev, R., Sabach, E., \& Soker, N. 2019, \apj, 884, 58

\bibitem[Shara et al.(1985)]{shara85}
Shara, M. M., Moffat, A. F. J., \& Webbink, R.F. 1985, \apj, 294, 271


\bibitem[Soker \& Tylenda(2003)]{soktyl}
Soker, N. \& Tylenda, R. 2003, \apj, 582, L105

\bibitem[Spada et al.(2017)]{Spada} 
Spada, F., Demarque, P., Kim, Y.-C., et al.\ 2017, \apj, 838, 161 

\bibitem[Sparks et al.(2008)]{sparks}
Sparks, W.~B., Bond, H.~E., Cracraft, M., et al.\ 2008, \aj, 135, 605

\bibitem[Staff et al.(2012)]{staff12}
Staff, J. E.; Menon, A.; Herwig, F. et al. 2012, \apj, 757, 76
  
\bibitem[Tylenda \& Soker(2006)]{tylsok}
Tylenda, R. \& Soker, N. 2006, \aap, 451, 223

\bibitem[Tylenda et al.(2011)]{thk11}
Tylenda, R., Hajduk, M., Kami\'nski, T. et al. 2011, \aap, 528, A114

\bibitem[Tylenda et al.(2013)]{tku13}
Tylenda, R., Kami\'nski, T., Udalski, A. et al. 2013, \aap, 555, A16

\bibitem[Tylenda et al.(2019)]{tk19}
Tylenda, R., Kami\'nski, T., \& Mehner, A., 2019, \aap, 628, A124

\bibitem[Young et al.(2001)]{tycho}
Young, P. A., Mamajek, E. E., Arnett, D., \& Liebert, J. 2001, \apj,
556, 230

\bibitem[Zamora et al.(2009)]{zamora}
Zamora, O., Abia, c., Plez, B. et al. 2009, \aap, 508, 909

\bibitem[Zhang \& Jeffery(2013)]{zj13}
Zhang, X., \& Jeffery, C. S., 2013, \mnras, 430, 2113

\bibitem[Zhang et al.(2020)]{zj20}
Zhang, X., Jeffery, C. S., Li, Y., \& Bi, S., 2020, \apj, 889, 33

\end{thebibliography}
\end{document}